\documentclass[aps,prl,reprint,showpacs,floatfix,amsmath,amssymb,10pt]{revtex4-1} 

\usepackage{graphicx}       
\usepackage{epstopdf}
\usepackage{txfonts}        

\usepackage[plainpages=false,pdfpagelabels,colorlinks]{hyperref} 
\hypersetup{colorlinks,citecolor=blue,filecolor=blue,linkcolor=blue,urlcolor=blue}

\begin{document}


\title{Determining the validity of the meanfield Bogoliubov-de Gennes equation}

\author{Brendan C. Mulkerin}%
\email{bmulkerin@pgrad.unimelb.edu.au}
\author{Andrew M. Martin}
\affiliation{School of Physics, University of Melbourne, Victoria 3010, Australia}

\date{\today}

\begin{abstract}

We provide a general methodology to directly determine the validity of the meanfield Bogoliubov-de Gennes equation. In particular we apply this methodology to the case of two component interacting ultracold Fermi gases. As an example, we consider the case of population imbalance, between the two components, in the strongly attractive interacting regime, where meanfield results predict Fulde-Ferrell-Larkin-Ovchinnikov (FFLO) states. For these states we find at finite temperatures that the assumptions used to derive the Bogoliubov-de Gennes equation are invalid.        
\end{abstract}

\maketitle


The onset of experimentally realized superfluidity of Fermionic gases by the MIT group in 2004 \cite{Ketterle:2002zz} has seen a trend towards theoretical studies of ultracold gases. With the use of a wide range of techniques at experimentalists fingertips, it is possible to tune ultracold gases to sit in specific potentials, have precise particle numbers and create mismatched Fermi surfaces giving rise to many new concepts to study and analogies to solids \cite{Bloch:2008zzb}. An important advancement was the use of Feshbach resonances in controlling the s-wave scattering length, $a_{s}$; this was the defining step in realizing superfluidity in two component fermionic systems. Applying an external magnetic field enables one to control the relative energy between the two channels of the scattering process: the open (scattering) and closed (bound) channels. This gives experimentalists the ability to form a smooth transition from the deep attractive to the strongly repulsed; the deep BCS side to tightly bound Bose-Einstein Condensation (BEC) interaction \cite{Bartenstein:2004,O'Hara2002,Chin2004,Kinast2005}.

Almost as soon as superfluidity was achieved in Fermi gases the question of how spin population imbalance would effect the system was asked. When a gas has a polarization $P=[N_{\uparrow}-N_{\downarrow}]/N$ greater than zero where $N$ is the total particle number, a difference in the chemical potential for each spin state is enforced by the system, which can be thought of equivalently as applying an external magnetic field. Building upon the work down by Bardeen, Cooper and Schrieffer \cite{Cooper:1956zz,Bardeen:1957mv}, the first analogous work done in superconductors was by Clogston and Chandresekaer \cite{Clogston:1962zz, chandrasekhar1962note} where they proposed a critical cut-off for the superconductivity in the presence of a magnetic field.

An original idea for imbalance in superconductors was proposed in the 1960's by Fulde and Ferrel \cite{Fulde:1964zz} and separately Larkin and Ovchinnikov \cite{larkin:1964zz}, they found a spin density imbalance in a system would induce non-zero momentum Cooper pairs leading to oscillations in the pairing field, the Fulde, Ferrel, Larkin and Ovchinnikov (FFLO) state. 
The experimental realization of population imbalance in Fermi gases  \cite{Zwierlein:2006,Partridge:1900zz,Partridge:2006zz,shin:2008} in 2006 caused a flurry of theoretical investigations into the FFLO state. These have primarily focused on meanfield Bogoliubov-de Gennes approaches \cite{Kinnunen:2006zz,Hui:2007,Machida:97,Ohashi:05,Perali:2004zz,jensen-2007}, which have been complemented by beyond meanfield Density Functional Theory \cite{Bulgac:2008tm}, variational techniques \cite{Sheehy:2006qc} and the T-matrix scheme \cite{He:2007ps} in the BEC-BCS crossover regime and 1D exact and Quantum Monte Carlo techniques \cite{zhao:2008,casula:2008}. This theoretical endeavor is complemented by experimental research \cite{schunck-2007-316, Ketterle:2009, Liao2010} to observe the FFLO state. 

In this work we derive a general methodology for determining the validity of the meanfield solutions of the Bogoliubov-de Gennes equation and implement this for the specific case of the well studied imbalanced ultracold Fermi gas in the strong interacting regime. 
As such, the paper  is set out in the following manner, firstly we  describe the Bogoliubov-de Gennes meanfield theory and its regimes of validity. Then we present results for a balanced gas and an imbalanced gas in a spherical trap, demonstrating, qualitatively, the regimes under which the meanfield solutions are valid. Finally, we comment on the validity of previous meanfield analysis of FFLO states in imbalanced ultracold Fermi gases.

Our starting point for the study of balanced and imbalanced ultracold Fermi gases is the following microscopic Hamiltonian
\begin{alignat}{1}
H= & \sum_{\sigma}\int\mathrm{d}^{3}\mathbf{r}\hat{\Psi}_{\sigma}^{\dagger}(\mathbf{r})\mathcal{H}_{\sigma}\hat{\Psi}_{\sigma}(\mathbf{r})\nonumber \\
 & +\int\mathrm{d}^{\mathrm{3}}\mathbf{r}\mathrm{d}^{3}\mathbf{r}'\hat{\Psi}_{\uparrow}^{\dagger}(\mathbf{r}')\hat{\Psi}_{\downarrow}^{\dagger}(\mathbf{r})U(\mathbf{r}-\mathbf{r}')\hat{\Psi}_{\downarrow}(\mathbf{r})\hat{\Psi}_{\uparrow}(\mathbf{r}')\label{eq:hamiltonian},\end{alignat}
where $\hat{\Psi}_{\sigma}^{\dagger}(\mathbf{r}),\ \hat{\Psi}_{\sigma}(\mathbf{r})$ are the real space annihilation and creation operators for an atom with spin $\sigma$ at position $\mathbf{r}$. $\mathcal{H}_{\sigma}=(-\hbar^{2}\nabla^{2}/2m_{\sigma}-\mu_{\sigma}+V_{\sigma}(\mathbf{r}))$ is the single particle Hamiltonian for a atom of mass $m_{\sigma}$ residing in a potential $V_{\sigma}(\mathbf{r})$, where $\mu_{\sigma}$ is the chemical potential for component $\sigma$. $U(\mathbf{r-r'})=U\delta(\mathbf{r-r'})$ is the contact interaction with $U=4\pi\hbar^2 a_s/m_r$, where $m_r$ is the reduced mass. 

Applying the meanfield approximation to Eq~\eqref{eq:hamiltonian} and defining the pairing field and spin density,  $\Delta(\mathbf{r})=U\langle\hat{\Psi}_{\downarrow}(\mathbf{r})\hat{\Psi}_{\uparrow}(\mathbf{r})\rangle $ and $n_{\sigma}(\mathbf{r})=\langle\hat{\Psi}_{\sigma}^{\dagger}(\mathbf{r})\hat{\Psi}_{\sigma}(\mathbf{r})\rangle $ respectively, we can write the meanfield Hamiltonian as \begin{alignat}{1}
H_{MF}= & \Omega_{G}+{\textstyle {\displaystyle \int\mathrm{d}^{3}\mathbf{r}}}\Bigr\{\sum_{\sigma}\hat{\Psi}_{\sigma}^{\dagger}(\mathbf{r})\mathcal{H}_{\sigma}\hat{\Psi}_{\sigma}(\mathbf{r})\nonumber\\
 & +\sum_{\sigma}U\hat{\Psi}_{\bar{\sigma}}^{\dagger}(\mathbf{r})n_{\sigma}(\mathbf{r})\hat{\Psi}_{\bar{\sigma}}(\mathbf{r})+\hat{\Psi}_{\uparrow}^{\dagger}(\mathbf{r})\Delta(\mathbf{r})\hat{\Psi}_{\downarrow}^{\dagger}(\mathbf{r})+H.c.\Bigl\}\label{eq:mean},\end{alignat} where $\Omega_G$ is the ground state energy.

It is well known that we can write Eq~\eqref{eq:mean} in a diagonalized form, $H_{MF}=\sum_{k\sigma} E_k \gamma^{\dagger}_{k\sigma}\gamma^{}_{k\sigma}$ by a Bogoliubov-Valatin transformation and solving the Bogoliubov-de Gennes equations \cite{degennes-66}
\begin{alignat}{1}\label{eq:BdG}
\left(\begin{array}{cc}
\mathcal{H}_{\uparrow} & \Delta(\mathbf{r})\\
\Delta^{\star}(\mathbf{r}) & -\mathcal{H}_{\downarrow}\end{array}\right)\left(\begin{array}{c}
u_{k}(\mathbf{r})\\
v_{k}(\mathbf{r})\end{array}\right)=E_{k}\left(\begin{array}{c}
u_{k}(\mathbf{r})\\
v_{k}(\mathbf{r})\end{array}\right),\end{alignat}
where the wavefunctions for a particle with spin $\sigma$, position $\mathbf{r}$ and time $t$ are given by
\begin{alignat}{1}
\Psi_\uparrow (\mathbf{r},t)=\sum_{k} \left[u_{k\uparrow}(\mathbf{r}) \gamma_{k\uparrow}e^{-iE_{k\uparrow}t/\hbar} -v_{k\downarrow}^{\star}(\mathbf{r})\gamma_{k\downarrow}^{\dagger}e^{iE_{k\downarrow}t/\hbar}\right] \\
\Psi_\downarrow ^{\dagger} (\mathbf{r},t)=\sum_{k} \left[u_{k\downarrow} ^ {\star} (\mathbf{r})\gamma_{k\downarrow} ^{\dagger} e^{iE_{k\downarrow}t/\hbar} +v_{k\uparrow}(\mathbf{r})\gamma_{k\uparrow} e^{-iE_{k\uparrow}t/\hbar}\right],
\end{alignat}
$\gamma^{}_{k\sigma}$ and $\gamma_{k\sigma}^\dagger$ are fermionic quasiparticle creation and annihilation operators and $u_{k}(\mathbf{r})$ and $v_{k}(\mathbf{r})$ are the Bogoliubov quasiparticle wave functions, with corresponding eigenenergies $E_k$.
We have introduced for completeness a Hartree term to the single particle Hamiltonian $\mathcal{H}_{\sigma}=-\hbar^{2}\nabla^{2}/2m_{\sigma}-\mu_{\sigma}+V_{\sigma}(\mathbf{r}))+Un_{\bar{\sigma}}(\mathbf{r})n_{\sigma}(\mathbf{r})$. The self-consistent pairing field $\Delta(\mathbf{r})$ and spin densities $n_{\sigma}(\mathbf{r})$ in terms of the quasiparticle wave functions are given by
\begin{alignat}{1}
n_{\uparrow}(\mathbf{r})= & \sum_{k}|u_{k}(\mathbf{r})|^{2}f(E_{k})\label{density up},\\
n_{\downarrow}(\mathbf{r})= & \sum_{k}|v_{k}(\mathbf{r})|^{2}f(-E_{k})\label{density down},\\
\Delta(\mathbf{r})= & \tilde{U}(\mathbf{r})\sum_{k}u_{k}(\mathbf{r})v_{k}^{\star}(\mathbf{r})f(E_{k})\label{delta},\end{alignat} where the Fermi distribution is $f(E_k)=[\exp(-E_k/k_{B}T)+1]^{-1}$ and $\tilde{U}(\mathbf{r})$ has been introduced to renormalize the ultraviolet divergences, discussed in detail later. These equations form the basis of most meanfield calculations of interacting ultracold Fermi gases, where the calculations are iterated to achieve self-consistency.

To move from Eqs.~\eqref{eq:hamiltonian} to \eqref{eq:mean} a meanfield approximation has been employed to the interacting part of the Hamiltonian and expanded about the small order parameters 
 \begin{alignat}{1}
\hat{\delta}({\mathbf r},t) & =\tilde{U}(\mathbf{r})\left[\hat{\Psi}_{\uparrow}(\mathbf{r},t)\hat{\Psi}_{\downarrow}(\mathbf{r},t)-\left\langle \hat{\Psi}_{\downarrow}(\mathbf{r},t)\hat{\Psi}_{\uparrow}(\mathbf{r},t)\right\rangle\right], \label{small delta} \\
\hat{\delta} n_{\sigma}({\mathbf r},t) & =\hat{\Psi}_{\sigma}^{\dagger}(\mathbf{r},t)\hat{\Psi}_{\sigma}(\mathbf{r},t)-\left\langle \hat{\Psi}_{\sigma}^{\dagger}(\mathbf{r},t)\hat{\Psi}_{\sigma}(\mathbf{r},t)\right\rangle, \end{alignat}
where we have assumed the second order terms of $\langle\hat{\delta}^{\dagger}({\mathbf r},t)\hat{\delta}({\mathbf r},t')\rangle$ and $\langle\hat{\delta} n_\sigma^{\dagger}({\mathbf r},t)\hat{\delta} n_\sigma({\mathbf r},t')\rangle$ go to zero. We can however calculate these terms retrospectively by defining the second order fluctuations in the pairing and density fields, for example in the pairing field
\begin{alignat}{1}
\langle\hat{\delta}^{\dagger}({\mathbf r},t)\hat{\delta}({\mathbf r},t')\rangle= & 
\left[\tilde{U}(\mathbf{r})\right]^2\biggl[\left\langle\hat{\Psi}^{\dagger}_{\uparrow}(\mathbf{r},t)\hat{\Psi}^{\dagger}_{\downarrow}(\mathbf{r},t)\hat{\Psi}_{\downarrow}(\mathbf{r},t')\hat{\Psi}_{\uparrow}(\mathbf{r},t')\right\rangle \nonumber \\ 
 &  -\left\langle\hat{\Psi}^{\dagger}_{\uparrow}(\mathbf{r},t)\hat{\Psi}^{\dagger}_{\downarrow}(\mathbf{r},t)\right\rangle\left\langle\hat{\Psi}_{\downarrow}(\mathbf{r},t')\hat{\Psi}_{\uparrow}(\mathbf{r},t')\right\rangle\biggl] \label{pairing fluctuations}.
\end{alignat}
In an analogous method for current/charge fluctuations in mesoscopic conductors \cite{buttiker:1992,Martin2000,B&uuml;ttiker2000}, the anomalous fluctuations are determined through the quantum statistical expectation value of the relation \cite{landaustatphys}
\begin{alignat}{1}
\langle\hat{\delta}^{\dagger}({\mathbf r},\omega)\hat{\delta}({\mathbf r},\omega')\rangle & 
=\left[\tilde{U}(\mathbf{r})\right]^2\biggl[\langle\hat{\Delta}({\mathbf r},\omega)\hat{\Delta}({\mathbf r},\omega')\rangle \nonumber  \\ 
& - \langle\hat{\Delta}({\mathbf r},\omega)\rangle\langle\hat{\Delta}({\mathbf r},\omega')\rangle\biggl] \label{eq:flucster omega},
\end{alignat}
where $\hat{\Delta}(\mathbf{r},\omega)=\hat{\Psi}_{\downarrow}(\mathbf{r},\omega)\hat{\Psi}_{\uparrow}(\mathbf{r},\omega)$ is the Fourier transform of the product of time space wavefunctions $\hat{\Delta}(\mathbf{r},t)=\hat{\Psi}_{\downarrow}(\mathbf{r},t)\hat{\Psi}_{\uparrow}(\mathbf{r},t)$ and similarly to Eq. (\ref{small delta}) we write
\begin{alignat}{1}
\hat{\delta}({\mathbf r},\omega) & =\tilde{U}(\mathbf{r})\biggl[\hat{\Delta}(\mathbf{r},\omega)-\langle\hat{\Delta}(\mathbf{r},\omega)\rangle\biggl] \label{fourier}.
\end{alignat}
Now we calculate the second order fluctuations in frequency space. Taking the Fourier transform of the product of wavefunctions using the Bogoliubov quasiparticle operators we have 
\begin{alignat}{1}
\hat{\Delta}(\mathbf{r},\omega)= &
2\pi\sum_{k}\Bigl[u_{\downarrow k}u_{\uparrow k+\hbar\omega}\gamma_{\downarrow k} \gamma_{\uparrow k+\hbar\omega}^{} +v^{\star}_{\uparrow k}u_{\uparrow k+\hbar\omega}\gamma^{\dagger}_{\uparrow k} \gamma_{\uparrow k+\hbar\omega} \nonumber \\
- &
u_{\downarrow k}^{ }v^{\star}_{\downarrow k-\hbar\omega}\gamma_{\downarrow k}^{ } \gamma^{\dagger}_{\downarrow k-\hbar\omega}-v^{\star}_{\uparrow k}v^{\star}_{\downarrow k-\hbar\omega}\gamma^{\dagger}_{\uparrow k} \gamma^{\dagger}_{\downarrow k-\hbar\omega}\Bigl]\label{transformed},
\end{alignat}
where the position $\mathbf{r}$ is implied; we then get a similar expression for $\hat{\Delta}^{\star}(\mathbf{r},\omega)$.
Using the identity (Eq. (1.15)) from \cite{buttiker:1992},
\begin{alignat}{1}
\langle\gamma_{\alpha m}^{\dagger}\gamma_{\beta n}^{}\gamma_{\gamma k}^{\dagger}\gamma_{\delta l}^{}\rangle-\langle\gamma_{\alpha m}^{\dagger}\gamma_{\beta n}^{}\rangle\langle\gamma_{\gamma k}^{\dagger}\gamma_{\delta l}^{}\rangle\nonumber \\
=\delta(E_{m}-E_{n})\delta(E_{k}-E_{l})\delta_{\alpha\delta}\delta_{\beta\gamma}f_{a}(E_{n})[1-f_{\gamma}(E_{m})],
\end{alignat}
with Eq. (\ref{transformed}) in Eq. (\ref{eq:flucster omega}) gives delta functions of the form $\delta(E_k-E_j-\hbar\omega')$ and $\delta(E_k+\hbar\omega-E_j)$, allowing us to perform one summation over the energies and set $\omega=-\omega'$, giving
\begin{widetext}
\begin{alignat}{1}
\langle\hat{\delta}^{\dagger}({\mathbf r})\hat{\delta}({\mathbf r})\rangle_{\omega} &
= \sum_k \left[ u^{\star}_{\uparrow k-\hbar\omega} v^{}_{\uparrow k} v^{\star}_{\uparrow k} u^{}_{\uparrow k-\hbar\omega} f(E_{\uparrow k-\hbar\omega})f(E_{\uparrow k}) + 
2u^{\star}_{\uparrow k-\hbar\omega} v^{}_{\uparrow k} v^{\star}_{\uparrow k} u^{}_{\uparrow k-\hbar\omega} f(E_{\uparrow k-\hbar\omega})(1-f(E_{\uparrow k})) \right. \nonumber \\ 
&\left.-u^{\star}_{\uparrow k+\hbar\omega} v^{}_{\uparrow k} v^{\star}_{\uparrow k} u^{}_{\uparrow k+\hbar\omega} f(E_{\uparrow k+\hbar\omega})f(E_{\uparrow k})+
2u^{\star}_{\uparrow k+\hbar\omega} v^{}_{\uparrow k} v^{\star}_{\uparrow k} u^{\star}_{\uparrow k+\hbar\omega}f(E_{\uparrow k+\hbar\omega})(1-f(E_{\uparrow k})) \right],
\end{alignat}
\end{widetext}
and in the limit $\omega \rightarrow 0$ we have the result,
\begin{alignat}{1}
\langle\hat{\delta}^{\dagger}({\mathbf r})\hat{\delta}({\mathbf r})\rangle & =8\pi^2\left[\tilde{U}(\mathbf{r})\right]^2\sum_{k}|u_{k}(\mathbf{r})|^{2}|v_{k}(\mathbf{r})|^{2}f(E_{k})f(-E_{k}).\label{eq:delta fluc}
\end{alignat}
We wish to look at the zero frequency limit as the pairing field, Eq. (\ref{delta})  (and densities Eq. (\ref{density up}, \ref{density down})) are calculated in the zero frequency limit.
We can calculate in a similar fashion to above and find the anomalous fluctuations in the densities of each spin state
\begin{alignat}{1}
\langle\delta n_{\uparrow}({\mathbf r})\delta n_{\uparrow}({\mathbf r})\rangle & =8\pi^2\sum_{k}|u_{k}(\mathbf{r})|^{4}f(E_{k})f(-E_{k}),\label{eq:density fluc1}\\
\langle\delta n_{\downarrow}({\mathbf r})\delta n_{\downarrow}({\mathbf r})\rangle & =8\pi^2\sum_{k}|v_{k}(\mathbf{r})|^{4}f(E_{k})f(-E_{k})\label{eq:density fluc2}.\end{alignat}  

In this paper we consider a spherically symmetric trap, i.e. $V({\mathbf r})=m_{\sigma}\omega^2 r^2/2$, with the atomic mass of each of the components being equal, i.e. $m_{\uparrow \downarrow}=m_{\uparrow}=m_{\downarrow}$. For such a case it is natural to expand the wave functions $\hat{\Psi}_{\sigma}(\mathbf{r})$ in the basis of the spherical harmonic oscillator eigenstates. In the usual quantum numbers, following Refs~\cite{Ohashi:05,Kinnunen:2006zz}, $\{k\}=\{n,m,l\}$ we have 
\begin{alignat}{1}
\hat{\Psi}_{\sigma}(r)=\sum_{nml}g_{nml}(r)\hat{a}_{nml\sigma},\end{alignat}
where $\hat{a}_{nml\sigma}$ are the fermionic operators, $g_{nml}(r)=R_{nl}(r)Y_{ml}(\theta,\phi)$ and
\begin{alignat}{1}
R_{nl}(r)=\sqrt{2}(m_{\uparrow\downarrow}\omega)^{3/4}\sqrt{\frac{n!}{(n+l+1/2)!}}\hat{r}^{l}L_{n}^{l+1/2}(\hat{r}^{2})e^{-\frac{\hat{r}^{2}}{2}},
\end{alignat}
are the normalized basis states, with $L_{n}^{l+1/2}(\hat{r}^{2})$ being a Laguerre Polynomial and $\hat{r}=r/a_{ho}$, with $a_{ho}=\sqrt{\hbar/m_{\uparrow \downarrow}\omega}$ being the harmonic oscillator length. As $m$ is a good quantum number we can calculate the angular integrations by introducing a $(2l+1)$ degeneracy to the $l$ states. The $l$-states of the Hamiltonian can be diagonalized separately due to the axial symmetry, greatly reducing the computational complexity of the problem. In order to diagonalize the Hamiltonian and calculate any quantities we must truncate the quasiparticle energies to $E_{nl}<E_{\mathrm{c}}$, where the choice of $E_{\mathrm{c}}$ is large enough to not change the final solution. Writing the order parameter and densities in this basis and introducing the $l$ specific cut-off $N_l=(E_{\mathrm{c}}-l-3/2)/2$ we obtain
 \begin{alignat}{1}
\Delta(r)=\tilde{U}(r)\sum_{nn'l}^{N_l}\frac{2l+1}{4\pi}R_{nl}(r)R_{n'l}(r){\langle}a_{nl\uparrow}^{\dagger}a_{n'l\downarrow}^{\dagger}\rangle, \label{eq:pair} \\
n_\sigma(r)=\sum_{nn'l}^{N_l}\frac{2l+1}{4\pi}R_{nl}(r)R_{n'l}(r){\langle}a_{nl\sigma}^{\dagger}a_{n'l\sigma}\rangle \label{eq:density}.
\end{alignat}
These self consistent equations are solved by iteration for fixed particle numbers $N_\uparrow$ (majority) and $N_\downarrow$ (minority) by adjusting the chemical potentials $\mu_\sigma$. In each iteration the Hamiltonian is diagonalized and the resultant eigenvectors and eigenenergies are used to calculate the pairing and number densities, Eqs. \eqref{eq:pair} and \eqref{eq:density}, until the relative change in the particle number and pairing field is  $\le 10^{-6}$ between consecutive iterations.

We turn our attention now to the problem of ultraviolet divergences in the pairing function and Hartree terms. The divergence of the Hartree terms, $Un_{\sigma}(\mathbf{r})$ is clear as we go to the unitary regime $a_{s}\rightarrow\infty$, then $U\rightarrow\infty$ making the Hartree terms infinitely small. We can overcome this divergence by eliminating the Hartree terms from the free Hamiltonian in the unitary regime, and in our calculations we have not included them \cite{Heiselberg:2001}. The divergence of the pairing field due to the nature of the contact interaction is easily solvable by regularizing the equation using the pseudo potential prescription \cite{Grasso:03}, with the inverse of the regularized contact interaction given by 
\begin{alignat}{1}
\frac{1}{\tilde{U}(\mathbf{r})}=&
\frac{1}{U}+\frac{m_{\uparrow \downarrow }}{2\pi^{2}\hbar^2}\left(\frac{k_{\mathrm{F}}(\mathbf{r})}{2} \mathrm{ln}\frac{k_{\mathrm{c}}(\mathbf{r})+k_{\mathrm{F}}(\mathbf{r})}{k_\mathrm{c}(\mathbf{r})-k_{\mathrm{F}}(\mathbf{r})}-k_{\mathrm{c}}(\mathbf{r})\right),
\end{alignat}
where the local Fermi momentum is given by  $\mu=\hbar^2k_{\mathrm{F}}(\mathbf{r})^2/2m_{\uparrow \downarrow}+V(\mathbf{r})$, $E_c=\hbar^2k_{\mathrm{c}}(\mathbf{r})/2m_{\uparrow \downarrow}+V(\mathbf{r})$ and $\mu=[\mu_{\uparrow}+\mu_{\downarrow}]/2$. 
Following from Ref. \cite{Hui:2007} we have used a hybrid scheme for calculating the gap equation, i.e. $\Delta(\mathbf{r})=\Delta^{\mathrm{BdG}}(\mathbf{r})+\Delta^{\mathrm{LDA}}(\mathbf{r})$, where $\Delta^{\mathrm{BdG}}(\mathbf{r})$ is given by Eq.~(\ref{eq:pair}) and $\Delta^{\mathrm{LDA}}(\mathbf{r})$ is calculated from the semi-classical solution to the pairing equation obtained from the local density approximation (LDA) for contributions above the cutoff $E_{\mathrm{c}}$. Specifically, 
\begin{alignat}{1}
\Delta^{\mathrm{LDA}}(\mathbf{r})= &  -U\int_{E_{c}}^{\infty}\mathrm{\frac{d^{3}\mathnormal{k}}{(2\pi)^{3}}}\left(\frac{\Delta(\mathbf{r})}{2E(\mathbf{r},k)}\left[1-f(E_{k\downarrow})-f(E_{k\uparrow})\right]\right.\nonumber \\
& 
\quad \left.-\frac{\Delta(\mathbf{r})}{2(\epsilon(\mathbf{r},k)-\mu)}\right)\label{eq:LDA},
\end{alignat} 
with $E_{\downarrow\uparrow}=E(\mathbf{r},k)\mp\delta\mu$, $\delta\mu=[\mu_{\uparrow}-\mu_{\downarrow}]/2$, $E(\mathbf{r},k)=\sqrt{(\epsilon(\mathbf{r},k)-\mu)^{2}+|\Delta(\mathbf{r})|^{2}}$
and $\epsilon(\mathbf{r},k)=k^{2}\hbar^2/2m_{\uparrow \downarrow}+V(r)$. Using the hybrid scheme we found the solutions converge for a lower cut-off, speeding up our calculations. We can also include the relative fluctuations from the higher order corrections to the LDA approximation, however, any additional relative fluctuations will be positive definite, making $\langle\delta^{\dagger}({\mathbf r})\delta({\mathbf r})\rangle$ larger. Thus Eq. \eqref{eq:delta fluc} provides a lower bound for $\langle\delta^{\dagger}({\mathbf r})\delta({\mathbf r})\rangle$.

Now that we have everything ready to begin solving the self consistent equations we can calculate our anomalous statistical averages, Eqs~\eqref{eq:delta fluc} and \eqref{eq:density fluc1}, in the harmonic oscillator basis. Writing the quasiparticle eigenvectors in terms of the eigenstates $W_{jn}^l$, $\alpha_{nl}(r)=\sum_{n}R_{nl}(r)W^{l}_{jn}$ and $\beta_{nl}(r)=\sum_{n}R_{nl}(r)W^{l}_{jn+N_l}$ we have
\begin{eqnarray}
& & \langle\delta^{\dagger}(\mathbf{r})\delta(\mathbf{r})\rangle =\left[\tilde{U}(\mathbf{r})\right]^2\sum_{jl}2\pi(2l+1)\alpha_{ml}(r)^{2}\beta_{nl}(r)^{2}f(E_{jl})f(-E_{jl})\label{eq:fluc1} \nonumber\\ & & \\  
& &\langle\delta n_{\downarrow}(r)\delta n_{\downarrow}(r)\rangle  =\sum_{jl}2\pi(2l+1)\alpha_{ml}(r)^{4}f(E_{jl})f(-E_{jl})\label{eq:fluc2}\\
& &\langle\delta n_{\uparrow}(r)\delta n_{\uparrow}(r)\rangle  =\sum_{jl}2\pi(2l+1)\beta_{ml}(r)^{4}f(E_{jl})f(-E_{jl})\label{eq:fluc3}.\end{eqnarray}The additional $(2l+1)$ comes from the degeneracy of the $l$ states when the initial sum over the $m$-quantum number was taken.

\begin{figure}[bt]
     \includegraphics{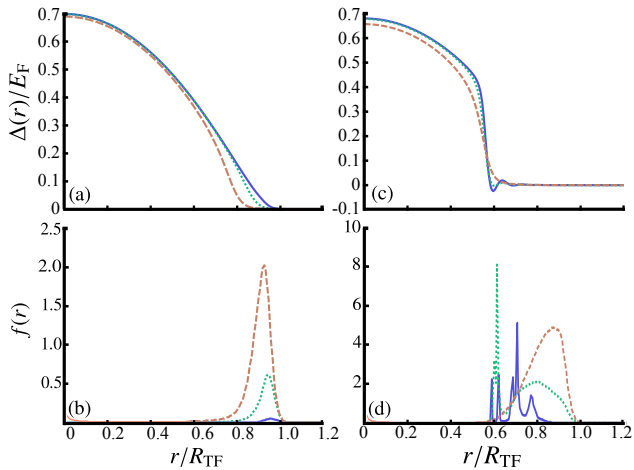}
  \caption{Spatial profiles of the (a,c) pairing field  [$\Delta(r)$] and the (b,d) the relative fluctuations in the pairing field [$f(r)=\sqrt{\langle\delta^{\dagger}(r)\delta(r)\rangle}/\Delta(r)$]. In each of the plots three temperatures are considered: $T=0.01T_\mathrm{F}$ (solid blue curve), $0.05T_\mathrm{F}$ (dotted green curve) and $0.1T_\mathrm{F}$ (dashed red curve) and the interaction strength is fixed at $U=-5$. For (a,b) $P=0$ and for (c,d) $P=0.5$.}
  \label{fig:pair and fluc}
\end{figure}

Since we are using 'trap units' the length and energy will be measured in harmonic oscillator length $a_{ho}=[\hbar/m_{\uparrow\downarrow}\omega]^{1/2}$ and $\hbar\omega$ respectively, and the temperature is in units of $T=\hbar\omega/k_\mathrm{B}$. Using the LDA prescription we obtain the Fermi Energy $E_\mathrm{F}=(3N)^{1/3}\hbar\omega$ and thus the Fermi temperature $T_\mathrm{F}=(3N)^{1/3}\hbar\omega/k_B$. We can also calculate the Thomas-Fermi radius $R_{\mathrm{TF}}=(24N)^{1/6}a_{ho}$. The calculations were performed for $N=10^5$ particles, at an interaction strength of $U=-5$ ($1/k_\mathrm{F}a_s=-0.22$) and a cut-off of $E_{c}=180\hbar\omega$ was chosen such that an increase in $E_{c}$ does not change the converged result. We have found that neglecting the Hartree terms does not qualitatively change the results presented here. Additionaly we only focus on the pairing field fluctuations, Eqs. \eqref{eq:delta fluc} and \eqref{eq:fluc1}, having found that the relative fluctuations in the densities Eqs. (\ref{eq:density fluc1}, \ref{eq:density fluc2}, \ref{eq:fluc2}, \ref{eq:fluc3}) do not give any additional information concerning the validity of the meanfield solution.

For a balanced system $(P=(N_{\uparrow}-N_{\downarrow})/N=0)$\ we regain the standard result for the pairing amplitude [Fig. \ref{fig:pair and fluc}(a)], i.e. away from the center of the trap $\Delta(r)$ decreases monotonically. The corresponding relative fluctuations in the pairing field, defined as $f(r)=\sqrt{\langle\delta^{\dagger}(r)\delta(r)\rangle}/\Delta(r)$, are plotted in Fig. \ref{fig:pair and fluc}(b). This plot shows that the relative fluctuations in the paring field, increase with temperature and become significant only when $\Delta(r)\rightarrow 0$. Since the Bogoliubov-de Gennes equation is valid for $f(r) \ll 1$ we can ascertain that the meanfield approximation used to obtain Fig. \ref{fig:pair and fluc}(a) is valid, except when $\Delta(r) \rightarrow 0$. 


For finite polaraization ($P=0.5$) the FFLO state manifests itself as an oscillation in the meanfield pairing amplitude, for $r/R_{\mathrm{TF}}>0.5$ [Fig. \ref{fig:pair and fluc}(c)]. This state emerges as the temperature is decreased. As for the balanced case we can investigate the validity of the meanfield approach through the determination of the relative fluctuations in the pairing amplitude [Fig. \ref{fig:pair and fluc}(d)]. Importantly we find that in the region the FFLO state emerges the meanfield approximation breaks down, i.e. $f(r) >1$, for  $T=0.01T_\mathrm{F}$ and $0.05T_\mathrm{F}$. However, we do note, as in the case of the balanced gas, that as the temperature is decreased the relative fluctuations in the pairing amplitude are reduced. As such, in this case, we expect at extremely low temperatures ($T<0.01T_\mathrm{F}$) that the meanfield predictions of FFLO states is valid.



\begin{figure}[bt]
     \includegraphics{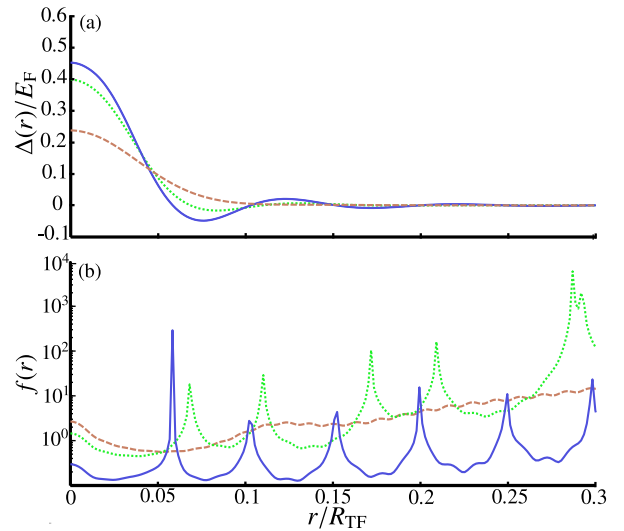}
  \caption{Spatial profiles of the (a) pairing field  [$\Delta(r)$] and the (b) relative fluctuations in the pairing field [$f(r)=\sqrt{\langle\delta^{\dagger}(r)\delta(r)\rangle}/\Delta(r)$]. In each of the plots three temperatures are considered: $T=0.01T_\mathrm{F}$ (solid blue curve), $0.05T_\mathrm{F}$ (dotted green curve) and $0.1T_\mathrm{F}$ (dashed red curve) with corresponding polarizations of $P=0.895$, $0.874$ and $0.841$. The interaction strength is fixed at $U=-5$}
  \label{fig:flucs at core}
\end{figure}

Ref. \cite{jensen-2007} considered the case where at a critical polarization the FFLO states extend to the core of the superfluid for zero temperature. We have performed similar calculations at finite temperatures ($T=0.01T_\mathrm{F}$, $0.05T_\mathrm{F}$ and $0.1T_\mathrm{F}$) in the strongly interacting regime of $U=-5$ to examine a polarized core, [Fig. \ref{fig:flucs at core}]. We found for polarizations of $P=0.895$ ( $0.874$) that at $T=0.01T_\mathrm{F}$ ($0.05T_\mathrm{F}$)  the FFLO state extends to the core of the superfluid [Fig. \ref{fig:flucs at core}(a)]. Additionally, for $T=0.1T_\mathrm{F}$ and $P=0.841$ there is no FFLO state [red dotted curve in Fig. \ref{fig:flucs at core}(a)], as calculated from the meanfield analysis. This is in stark contrast to the zero temperature calculation \cite{jensen-2007}, where an FFLO state extending to the core is predicted. For $T=0.1T_\mathrm{F}$ ($P=0.841$) the relative fluctuations about the meanfield ($f(r)$) are of order one throughout the superfluid [Fig. \ref{fig:flucs at core}(b)]. As such it is difficult to justify the validity of this meanfield solution. For the temperatures $T=0.05T_\mathrm{F}$ and $0.01T_\mathrm{F}$, where the meanfield FFLO state exists, the average of $f(r)$ over the range $r<0.15 R_\mathrm{TF}$ is greater than 1. As in the case of the edge FFLO states, shown in Fig. \ref{fig:pair and fluc}, the relative fluctuations about the meanfield decrease as the temperature is reduced. Specifically, for the parameters considered calculations show that the meanfield analysis only becomes valid for $T<0.001T_\mathrm{F}$,  an experimentally {\it challenging} regime. 

Overall we have provided a general formalism for determining the validity of meanfield Bogoliubov-de Gennes solutions based on the {\it retrospective} determination of the fluctuations about the meanfield, $f(r)$. We have used this method to show specifically, the validity of the FFLO state in a spherically symmetric ultracold gas. The analysis of the meanfield approximation at finite temperatures and zero polarization shows the fluctuations to be small and confined to the boundary of the superfluid, i.e. as $\Delta(r)\rightarrow 0$, indicating the meanfield approximation to be valid. However for a non-zero polarization we have seen $f(r)$ at the boundary become more pronounced and the relative fluctuations become significant, casting doubt on the meanfield results. When we examine the FFLO states at the core of the superfluid, the relative fluctuations extend with the modulating pairing field and are significant for temperatures lower than $T=0.01T_{\mathrm{F}}$. Only when we go several orders of magnitude below the current experimental temperatures can one be convinced of the meanfield results for finite polarization. In this study we have looked at a spherically symmetric trap, however one could extend this to asymmetric systems to consider the role of geometry and the validity of the meanfield approximation. 

\bibliography{Paper2}

\end{document}